\documentclass{article}
\usepackage{spconf,amssymb,amsmath,graphicx,bm,pifont}
\usepackage{tikz,multirow,pgfplots,subcaption,setspace}
\title{Mixup-Breakdown: A Consistency Training Method for\\Improving Generalization of Speech Separation Models}
\name{Max W. Y. Lam$^{\star}$ \qquad Jun Wang$^{\star}$ \qquad Dan Su$^{\star}$ \qquad Dong Yu$^{\dagger}$}
\address{$^{\star}$ Tencent AI Lab, Shenzhen, China\\$^{\dagger}$ Tencent AI Lab, Bellevue WA, USA} 
\begin{document}
\ninept
\maketitle
\begin{abstract}
\small
Deep-learning based speech separation models confront poor generalization problem that even the state-of-the-art models could abruptly fail when evaluating them in mismatch conditions. To address this problem, we propose an easy-to-implement yet effective consistency based semi-supervised learning (SSL) approach, namely Mixup-Breakdown training (MBT). It learns a teacher model to ``breakdown" unlabeled inputs, and the estimated separations are interpolated to produce more useful pseudo ``mixup" input-output pairs, on which the consistency regularization could apply for learning a student model. In our experiment, we evaluate MBT under various conditions with ascending degrees of mismatch, including unseen interfering speech, noise, and music, and compare MBT's generalization capability against state-of-the-art supervised learning and SSL approaches. The result indicates that MBT significantly outperforms several strong baselines with up to 13.77\% relative SI-SNRi improvement. Moreover, MBT only adds negligible computational overhead to standard training schemes. 

%Recent advances of deep-learning-based speech separation models have drastically advanced the state-of-the-art performances on several benchmark datasets. However, when these models are asked to separate mixture signals that contain mismatch interference during inference against training, even the state-of-the-art model could fail abruptly. We attribute the problem of poor generalization to the limitation of existing training strategies, where some clean sources are artificially mixed into speech signals to produce labeled data. In such a supervised framework, training a model that generalizes well , which, however, are difficult to collect. Alternatively, this paper proposes an easy-to-implement yet effective semi-supervised training approach, namely, \textit{Mixup-Breakdown} training (MBT), which is a kind of semi-supervised training strategy designed for speech separation. In MBT, a teacher model is introduced to generate pseudo-labels, known as \textit{breakdown} pairs, from unlabeled mixture signals with a random SNR mixing scheme. The \textit{breakdown} pairs are then used to maintain the prediction consistency between the teacher and student models. For testing, we evaluate the performance of separation models when dealing with speech mixtures containing unseen interference. The result suggests that MBT outperforms state-of-the-art semi-supervised training techniques in speech separation, while significantly improving generalization of speech separation models.
\end{abstract}

\section{Introduction}
\label{sec:intro}

%Traditionally, there is a fundamental assumption in many speech processing algorithms that input signals are relatively clean or potentially corrupted by some simple noise, such as additive Gaussian noise. Yet, in real-world scenarios such an assumption always become problematic since speech signals recorded in the nature mostly are the complex mixtures that contain non-stationary interference, for example, environmental noise, music and speech from other talkers. Directly feeding the mixture signals into systems that are only trained on clean signals could lead to an abrupt drop in performance \cite{seltzer2013investigation, narayanan2014investigation}. To tackle this problem, speech separation has been practically used as a front-end module for these systems \cite{Du2014RobustSR, gao2015joint, wang2016joint, lam2019extract}.
\par
%Speech separation (a.k.a the ``cocktail party problem'' \cite{cherry1953some}) aims at extracting a relatively clean speech signal from a mixture signal. 
%With speech as the most direct means of human communication, humans could almost effortlessly conduct speech separation by consciously following the target speaker's voice even in the midst of noisy throng of party goers. Although recent advances of deep learning methods have been dramatically advanced the performance of speech separation models, building an automatic speech separation system that matches human-level performance remains challenging to date. A vital distinction of current speech separation systems from human auditory systems appears to be the ignorance of the unlabeled mixture signals in training. While most signals we perceived are naturally mixtures of different kinds of sound sources, it is almost impossible for us to listen to the clean version of each source before being able to distinguish it from the others, even if it is a new type of sound unseen previously.
\par
Recent advances of deep-learning-based speech separation models have drastically advanced the state-of-the-art performances on several benchmark datasets. Typical successful models include the high-dimensional-embedding based methods proposed initially as a deep clustering network (DPCL) \cite{isik2016single}, their extensions such as deep attractor network (DANet) \cite{chen2017deep}, deep extractor network (DENet) \cite{wang2018deep}, and anchored DANet (ADANet) \cite{luo2018speaker}, and also include permutation-invariant-training (PIT) based methods \cite{xu2018single, yu2017permutation, kolbaek2017multitalker}, which determine the correct output permutation by calculating the lowest value on an objective function through all possible output permutations, as well as the recently proposed Conv-TasNet \cite{luo2018conv}, which is a fully-convolutional time-domain network trained with PIT.

However, when evaluating these models with mixture signals that contain mismatch interference during inference against training, even the cutting-edge model could abruptly fail \cite{luo2018conv}. Essentially, training a large number of parameters in a complex neural network that generalizes well requires a large-scale, wide-ranging, sufficiently varied training data. On the one hand, collecting high-quality labeled data for speech separation is often expensive, onerous, and sometimes impossible; although augmenting the labeled data \cite{zhang2017mixup} can empirically improve the generalization of the models, we argue that the improvement is limited as no extra new input information could be exploited. On the other hand, unlabeled data, i.e., mixture signals, usually are voluminous and easy to acquire, yet, unfortunately, lack effective methods to make use of them and thus are ignored by most conventional deep-learning-based speech separation systems. 
Therefore, it is desirable to exploit unlabeled data effectively. This issue has been widely studied in semi-supervised learning (SSL) domain \cite{chapelle2009semi, rasmus2015semi}, and consistency-based methods \cite{sajjadi2016regularization, laine2016temporal, tarvainen2017mean, miyato2018virtual, luo2018smooth, verma2019interpolation} are one of the most promising research directions in SSL. Their fundamental assumption is the invariance of predictions under perturbations or transformations \cite{sajjadi2016regularization}, which leads to better generalization for the unexplored areas where unlabeled inputs lie on.
%On the contrary, conventional deep-learning-based speech separation systems are all trained on labeled mixture signals. Despite their remarkable performances on the benchmark datasets \cite{luo2018conv}, when given mixture signals that contain interference unseen in training, even the most promising system could fail abruptly. We argue that, in a supervised learning setting, training a separation model to generalize well essentially requires large-scale, sufficiently varied and relatively clean source signals, which, however, are expensive to collect. In fact, mixture signals are overwhelming in the nature, therefore it is preferable to make use of unlabeled mixture signals for improving the generalization ability of speech separation models. 

In particular, this paper presents a novel, useful, and easy-to-implement consistency based SSL algorithm, namely Mixup-Breakdown training (MBT), for speech separation tasks. In MBT, a mean-teacher model is introduced to predict separation outputs from input mixture signals, and especially unlabeled ones; these intermediate outputs (namely \textit{Breakdown}) apply with random interpolation mixing scheme, and then treated as fake ``labeled" mixture (namely \textit{Mixup}) to update the student model by minimizing the prediction consistency between the teacher and student models. In the experiment, we evaluate the performance of MBT models when dealing with speech mixtures containing unseen interference. The result suggests that MBT outperforms cutting-edge SSL training techniques \cite{tarvainen2017mean, verma2019interpolation} for speech separation tasks, and significantly improves generalization of the speech separation model.
\par
To the best of our knowledge, this paper is the first work that applies semi-supervised learning to speech separation with evidence of enhanced generalization over mismatch interference. The rest of this paper organizes as follows: Section \ref{sec:2} describes the main contribution of this paper -- our proposed Mixup-Breakdown training methodology after formalizing the conventional training practice in speech separation as well as discussing its intrinsic problem with generalization; Section \ref{sec:3} briefly reviews related works in the prior art; Section \ref{sec:4} describes the experimental setup and then studies the performance of separation models trained with different approaches under various kinds of unseen interference; Section \ref{sec:conc} finally concludes our work.
\vspace{-0.5em}
\section{Proposed Approach}
\vspace{-0.5em}
\label{sec:2}
%Recent success of deep learning models mainly stems from its capability of modeling abstractions on high-dimensional manifolds, which in turns requires learning a large number of parameters. 
%\iffalse
\begin{figure*}[t]
    \vspace{-0.68em}
    \centering
    \includegraphics[scale=0.4]{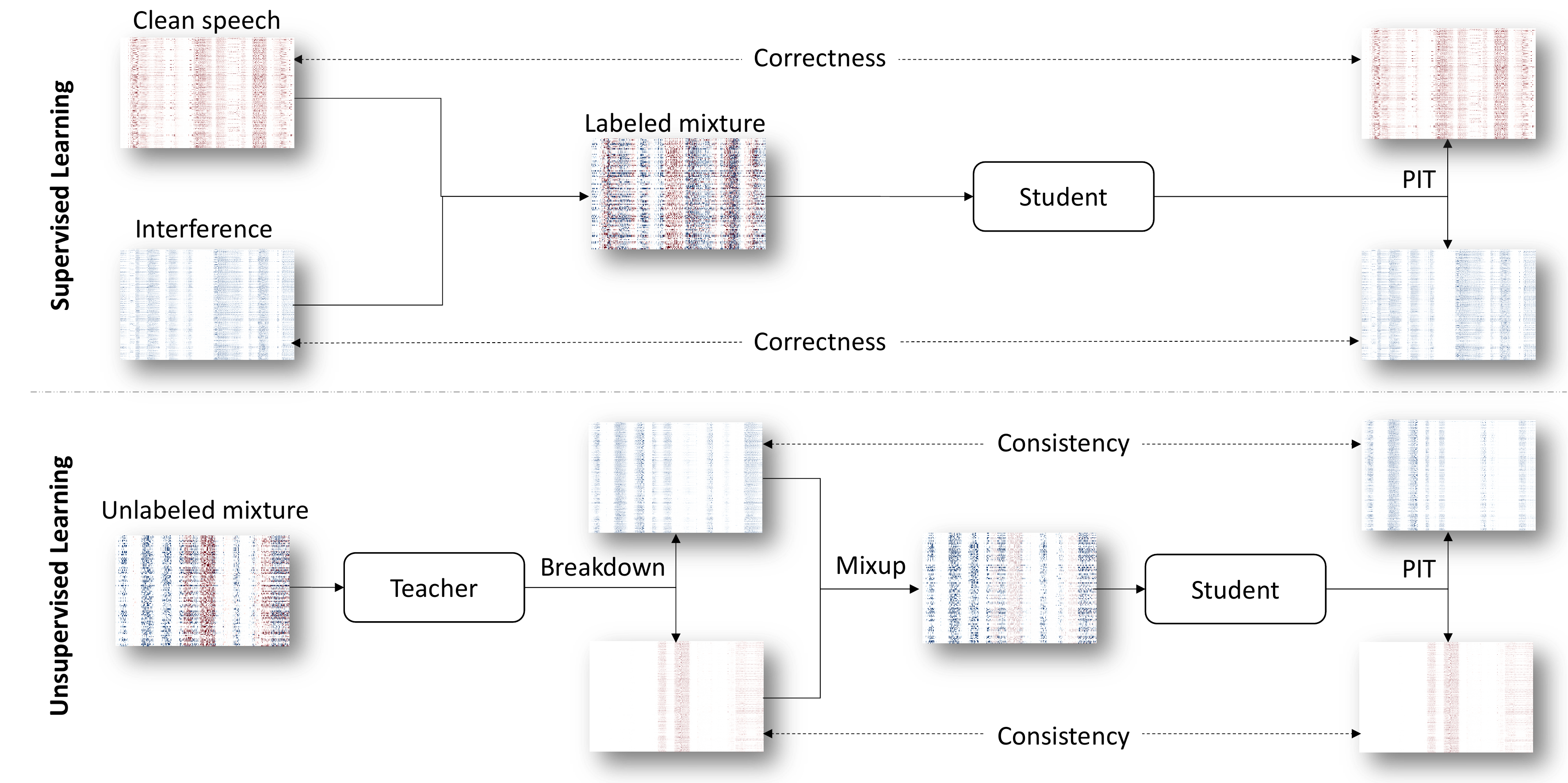}  %0.35
    \vspace{-0.68em}
    \caption{\small Algorithmic flow of the Mixup-Breakdown training divided into the supervised and unsupervised learning procedures}
    \label{fig:1}
    \vspace{-1.8em}
\end{figure*}
%\fi
Since large-scale, sufficiently varied training data with labels are often unattainable due to time and financial constraints, training a deep network of high complexity with a large number of parameters is prone to over-fitting and poor generalization. Therefore, we study a novel semi-supervised learning method called Mixup-Breakdown training (MBT) that exploits unlabeled data effectively to enhance generalization and reduce over-fitting.

As a typical training setup in speech separation, pairs of clean speech signal $\mathbf{s}$ and interference signal $\mathbf{e}$ add up into mixture signals $\mathbf{x}$ following randomly chosen signal-to-noise ratio (SNR) within specified range, resulting in a labeled training set of $N_L$ input-output pairs $\mathcal{D}_L=\{\mathbf{x}_i, \mathbf{y}_i\}_{i=1}^{N_L}$, where $\mathbf{y}=(\mathbf{s}, \mathbf{e})$, $\mathbf{x}=\mathbf{s}+ \mathbf{e}$. Aside from the labeled data, in practice, unlabeled data $\mathcal{D}_U=\{\mathbf{x}_j\}_{j=N_L+1}^{N=N_L+N_U}$ are usually more authentic, attainable, yet under-explored.

In a supervised learning framework, given a speech separation model $\mathbf{f}_\theta$ with parameters $\theta$, an objective function $\mathcal{L}(\mathbf{f}_\theta(\mathbf{x}), \mathbf{y})$ is usually defined as the divergence between the predicted outputs $\mathbf{f}_\theta(\mathbf{x})=(\mathbf{\hat{s}}, \mathbf{\hat{e}})$ and the original clean sources $\mathbf{y}$ to measure the ``correctness of separation'', as shown in the upper algorithmic flow in Fig. \ref{fig:1}. For example, we apply the recently proposed scale-invariant signal-to-noise ratio (SI-SNR) \cite{luo2018conv, yang2019improved} with PIT \cite{yu2017permutation, kolbaek2017multitalker}:
\begin{align}
    \mathcal{L}(\mathbf{f}_\theta(\mathbf{x}), \mathbf{y})&=\min _{\mathbf{u}\in\{\mathbf{\hat s}, \,\mathbf{\hat e}\}}\mathcal{L}_\text{SI-SNR}(\mathbf{s},\mathbf{u})+\min_{\mathbf{v}\in\{\mathbf{\hat s}, \,\mathbf{\hat e}\}}\mathcal{L}_\text{SI-SNR}(\mathbf{e},\mathbf{v}),\\
    \mathcal{L}_\text{SI-SNR}(\mathbf{a},\mathbf{b})&=-10\log_{10}\frac{\lVert\Pi_{\mathbf{a}}(\mathbf{b})\rVert_2^2}{\lVert\mathbf{b}-\Pi_{\mathbf{a}}(\mathbf{b})\rVert_2^2},
\end{align}
where
$
    \Pi_{\mathbf{a}}(\mathbf{b})=\mathbf{a}^\top\mathbf{b}/\lVert\mathbf{a}\rVert_2^2\cdot\mathbf{a}
$ is a projection of $\mathbf{b}$ onto $\mathbf{a}$.
\vspace{-0.5em}
\subsection{Conventional Supervised Learning}
\vspace{-0.3em}
We first formalize the conventional supervised learning framework. Assuming that the input-output pairs follow a joint distribution $P(\mathbf{x}, \mathbf{y})$, which is usually unknown, we minimize the average of the objective function over the joint distribution, i.e., the expected risk, to find an optimal set of parameters $\theta^*$:
\begin{align}
    \theta^*
    % &=\mathop{\arg\min}_{\theta}\int \mathcal{L}(\mathbf{f}_\theta(\mathbf{x}), \mathbf{y})\, d\,P(\mathbf{x}, \mathbf{y})\label{eq:1}\\
    &\approx\mathop{\arg\min}_{\theta}\int \mathcal{L}(\mathbf{f}_\theta(\mathbf{x}), \mathbf{y})\, d\,P_\text{EMP}(\mathbf{x}, \mathbf{y};\mathcal{D}_L)\label{eq:2}\\
    &=\mathop{\arg\min}_{\theta}\frac{1}{N_L}\sum_{i=1}^{N_L} \mathcal{L}(\mathbf{f}_\theta(\mathbf{x}_i), \mathbf{y}_i),\label{eq:3}
\end{align}
where in Eq.\ref{eq:2}, to approximate the unknown joint data distribution $P(\mathbf{x}, \mathbf{y})$, an empirical distribution is used:
\begin{align}
d\,P_\text{EMP}(\mathbf{x}, \mathbf{y};\mathcal{D}_L)=\frac{1}{N_L}\sum_{i=1}^{N_L}\delta(\mathbf{x}=\mathbf{x}_i, \mathbf{y}=\mathbf{y}_i)
\end{align}
where $\delta(\cdot)$ is a Dirac mass centered at $(\mathbf{x}_i, \mathbf{y}_i)$, so that the expected risk can estimate from the $N_L$ labeled training examples. 

This conventional approach is also known as the empirical risk minimization (ERM) \cite{vapnik1998statistical}. 
However, as highlighted in recent research \cite{zhang2017understanding, zhang2017mixup} as well as in classical learning theory \cite{vapnik1971uniformconv}, ERM has intrinsic limitations that the large neural networks trained with it memorize (instead of generalizing from) the training data; moreover, they are vulnerable to adversarial attacks, as they prone to produce drastically different predictions when giving examples just outside the training distribution. This evidence suggests that ERM is unable to generalize a model on testing distributions that differ only slightly from the training data. Our experiment in Section \ref{sec:generalization} also echoes this conclusion in mismatch training and test scenarios.
\vspace{-0.5em}
\subsection{Mixup-Breakdown}
\vspace{-0.3em}
The Mixup-Breakdown strategy is inspired from the fact that a human auditory system is capable of separating sources, not requiring any clean source for learning, yet having high consistency to perturbations such as high or low energies, fast or slow articulating speeds, moving or static locations, and with or without distortions. 

As illustrated in the algorithmic flow of MBT in Fig. \ref{fig:1}, we explore the interpolations between separated signals to provide a perturbation strategy to maintain the consistency of learning. 
We first introduce the Mixup and Breakdown operations:
\begin{align}
    &\text{Mix}_\lambda(\mathbf{a}, \mathbf{b}) \triangleq \lambda\cdot \mathbf{a} + (1-\lambda)\cdot \mathbf{b} \label{eq:mix}\\
    &\text{Break}_\lambda(\mathbf{a}, \mathbf{b}) \triangleq (\lambda\cdot \mathbf{a}, \,\,(1-\lambda)\cdot \mathbf{b})\label{eq:break}
\end{align}
where $\mathbf{a}$ and $\mathbf{b}$ are two arbitrary signals, and $\lambda\sim \text{Beta}(\alpha, \alpha)$ for $\alpha\in(0,\infty)$ is inherited from the \textit{mixup} approach \cite{zhang2017mixup}.

The Mixup-Breakdown (MB) strategy trains a student model $\mathbf{f}_{\theta_S}$ to provide consistent predictions with the teacher model $\mathbf{f}_{\theta_T}$ of the same network structure at perturbations of predicted separations from the input mixtures (either labeled or unlabeled):
\begin{align}
    \mathbf{f}_{\theta_S}(\text{Mix}_{\lambda}(\mathbf{f}_{\theta_T}(\mathbf{x}_j)))\approx \text{Break}_{\lambda}(\mathbf{f}_{\theta_T}(\mathbf{x}_j))
\end{align}
where the teacher model parameters $\theta_T$ is an exponential moving average of the student model parameters $\theta_S$. It has been proven that averaging model parameters over training steps tends to generate a more accurate model than directly using the final parameters \cite{polyak1992accel, tarvainen2017mean}, resulting in better generalization performance.

Meanwhile, adding perturbation to the estimated separations is likely to construct more useful pseudo labeled input-output pairs nearer to the separation boundary, on which the consistency regularization should apply.
Mathematically, the MB operation can view as a generic augmentation of the empirical distribution:
\begin{align}
    &d\,P_\text{MBT}(\mathbf{\tilde{x}}, \mathbf{\tilde{y}};\mathcal{D})=\frac{1}{N}\sum_{i=1}^N \bm{v}(\mathbf{\tilde{x}}, \mathbf{\tilde{y}}|\mathbf{x}_i)\\
    &\bm{v}(\mathbf{\tilde{x}}, \mathbf{\tilde{y}}|\mathbf{x}_i)=\mathbb{E}_{\lambda}\left[\delta\left(\mathbf{\tilde{x}}=\text{Mix}_{\lambda}(\mathbf{f}_{\theta_T}(\mathbf{x}_i)),\mathbf{\tilde{y}}=\text{Break}_{\lambda}(\mathbf{f}_{\theta_T}(\mathbf{x}_i))\right)\right]
\end{align}
% Essentially, the MB operation can be easily implemented with a minimal computation overhead.

%Essentially, we observe that the definition of input-output pairs in speech separation tasks is peculiar in a way that the outputs $(\mathbf{s}, \mathbf{e})$ can be interpolated to generate different valid inputs with different interpolation weights, and in fact this also is how we simulate datasets in practice. Inspired by this observation plus the recently proposed \textit{mixup} \cite{zhang2017mixup} operation, we propose a Mixup-Breakdown strategy to generate the pseudo online input-output pairs during training rather than using fixed SNRs defined beforehand. 

%where the \textit{breakdown} data pairs $(\bar{\mathbf{x}}_{j}, \bar{\mathbf{y}}_{j})$ are treated similarly as for the labeled pairs $(\mathbf{x}_{j}, {\mathbf{y}}_{j})$ that a weighted PIT-based SI-SNRi loss is used to measure the consistency leveraging the unlabeled input $\mathbf{x}_j$. 
%Note that $r(t)$ is necessary in MBT to determine the reliability of the Mixup-Breakdown, therefore it should be set to monotonically increase with the training iterations.

In a semi-supervised learning setting, provided the dataset $\mathcal{D}$ composed of the labeled dataset $\mathcal{D}_L$ and the unlabeled dataset $\mathcal{D}_U$,
we present a new consistency-based training method, namely, Mixup-Breakdown Training (MBT):
\begingroup
\allowdisplaybreaks
\begin{align}
    \theta^*_S\approx\mathop{\arg\min}_{\theta_S}&\underbrace{\left[\int \mathcal{L}(\mathbf{f}_{\theta_S}(\mathbf{x}), \mathbf{y})\, d\,P_\text{EMP}(\mathbf{x}, \mathbf{y};\mathcal{D}_L)\right.}_\text{Correctness}+\label{eq:2_1}\\
    &r(t)\underbrace{\left.\int\mathcal{L}(\mathbf{f}_{\theta_S}(\mathbf{\tilde{x}}), \mathbf{\tilde{y}})\, d\,P_\text{MBT}(\mathbf{\tilde{x}}, \mathbf{\tilde{y}};\mathcal{D})\right]}_\text{Consistency}\label{eq:2_1}\\
    =\mathop{\arg\min}_{\theta_S}&\left[\frac{1}{N_L}\sum_{i=1}^{N_L} \mathcal{L}(\mathbf{f}_{\theta_S}(\mathbf{x}_i), \mathbf{y}_i)+\right.\\
    &\left.\frac{r(t)}{N}\sum_{j=1}^{N}\mathcal{L}(\mathbf{f}_{\theta_S}(\text{Mix}_{\lambda}(\mathbf{f}_{\theta_T}(\mathbf{x}_j))),\,\, \text{Break}_{\lambda}(\mathbf{f}_{\theta_T}(\mathbf{x}_j)))\right],\label{eq:loss_mbt}
\end{align}
\endgroup
where $r(t)$ is the ramp function that increases the importance of the consistency term as the training goes \cite{tarvainen2017mean}. 
%As reported by the author \cite{zhang2017mixup}, the \textit{mixup} can improve the model performance and robustness in both image classification and speech recognition tasks.

% However, for unlabeled mixture signals $\mathbf{x}_j\in\mathcal{D}_U$, the individual sources are unknown. Therefore, we use the outputs of the mean-teacher model, denoted as the \textit{breakdown} pairs $(\bar{\mathbf{s}}_{j}, \bar{\mathbf{e}}_{j})=\mathbf{f}_{\theta_T}(\mathbf{x}_j)$, to create the \textit{mixup} pairs $(\bar{\mathbf{x}}_{j}, \bar{\mathbf{y}}_{j})$:
% \begin{align}
%     \bar{\mathbf{x}}_{j}\triangleq\mathcal{I}_\lambda(\bar{\mathbf{s}}_j, \bar{\mathbf{e}}_j),\,\,\,\,\bar{\mathbf{y}}_{j}\triangleq(\lambda\bar{\mathbf{s}}_j, (1-\lambda)\bar{\mathbf{e}}_j),
% \end{align}
% where the setting of $\lambda\sim \text{Beta}(\alpha, \alpha)$ is inherited from the \textit{mixup} approach. 
\subsubsection{Data Augmentation Effect Using Unlabeled Data}
\label{sec:3}
Data augmentation is a widely adopted technique to improve the generalization of a supervised model. For example, in image classification, new images are produced by shifting, zooming in/out, rotating, or flipping images \cite{bjerrum2017smiles}; likewise, in speech recognition, the training data are augmented using varied vocal tract length \cite{jaitly2013vocal}, SNR, tempo, and speed perturbation \cite{ko2015audio}, etc. However, these approaches mostly confine to labeled data. Notably, some recent research, including the \textit{mixup} technology \cite{zhang2017mixup} and generative adversarial networks (GANs), has also contributed to generic data augmentation methods \cite{goodfellow2014generative,antoniou2017data,verma2019interpolation}.

Unlike GANs that require training an additional model and increasing the complexity, our MBT can be easily implemented with minimal computational overhead. By observing Eq. \ref{eq:loss_mbt}, we can use MBT to manipulate both labeled data (i.e., for $j\in \{1, ..., N_L\}$) and unlabeled data (i.e., for $j\in \{N_L+1, ..., N\}$) to produce pseudo ``labeled'' input-output pairs outside the empirical distribution. Although, in this paper we focus on amplitude interpolation (Eq. \ref{eq:mix} and \ref{eq:break}) like different SNR augmentation, the MBT methodology is straightforward to extend to other types of perturbations, such as various distortions, speeds, and locations (for multi-channel scenarios). Therefore we consider that the proposed MBT is promising and may enlighten a new generic way to exploit unlabeled data.

\vspace{-0.5em}
\section{Related Work}
\vspace{-0.5em}
\label{sec:3}
Among all consistency-based methods, the recent interpolation consistency training (ICT) \cite{verma2019interpolation} 
have achieved the state-of-the-art results in computer vision (CV) benchmarks. The main difference of the ICT from the MT \cite{tarvainen2017mean} is the use of the aforementioned \textit{mixup} technique when calculating the consistency loss:
\begin{align}
    &\mathcal{L}_\text{ICT}=\mathcal{L}(\mathbf{y}_i, \mathbf{f}_{\theta_S}(\mathbf{x}_i))+r(t)\mathcal{C}(\mathbf{x}_{j} ,\mathbf{x}_{k})\\
    &\mathcal{C}(\mathbf{x}_{j} ,\mathbf{x}_{k})=\left\Vert \mathbf{f}_{\theta _{S}}\left(\text{Mix}_{\lambda}(\mathbf{x}_{j} ,\mathbf{x}_{k})\right) -\text{Mix}_{\lambda}(\mathbf{f}_{\theta _{T}}(\mathbf{x}_{j}) ,\mathbf{f}_{\theta _{T}}(\mathbf{x}_{k})) \ \right\Vert ^{2}_{2}
\end{align}
for $(\mathbf{x}_i, \mathbf{y}_i)\sim\mathcal{D}_L, \mathbf{x}_j,\mathbf{x}_k\sim\mathcal{D}_U$. Note that the \textit{mixup} here is on two input samples randomly drawn from the unlabeled data, which is fundamentally different from MBT.

As reported by the authors, ICT outperforms other cutting-edge SSL methods, including MT, with high significance in benchmark CV datasets. Therefore, we will analyze its performance as another strong baseline. Meanwhile, we use ICT as an ablation study to validate the necessity of the ``Breakdown" part in our   ``Mixup-Breakdown" methodology.

\vspace{-0.5em}
\section{Experiments}
\vspace{-0.5em}
\label{sec:4}
\subsection{Experimental Setup}
\vspace{-0.3em}
\subsubsection{Data Preparation}
\vspace{-0.2em}
To test our hypothesis that conventional models could abruptly fail when separating mixture signals that contain mismatch interference between test and training, we built three new datasets of speech mixtures based on the WSJ0-2mix corpus \cite{hershey2016deep} by replacing the background speech (the one with the lower SNR) with other types of interference :
\begin{itemize}
    \item WSJ0-Libri: using clean speech drawn from the publicly available Librispeech 100h training corpus \cite{panayotov2015librispeech}.
    \item WSJ0-music: using music clips drawn from a 43-hour music dataset that contains various classical and popular music genres, e.g., baroque, classical, romantic, jazz, country, and hip-hop.
    \item WSJ0-noise: using noise clips drawn from a 4-hour recording collected in various daily life scenarios such as office, restaurant, supermarket, and construction place.
\end{itemize}
Note that each of the above new datasets follows the same SNR range as WSJ0-2mix, and contains non-overlapping train, dev, and test set like WSJ0-2mix. We also combined all  the above unlabeled datasets into one union unlabeled dataset denoted as WSJ0-multi.
\vspace{-0.5em}
\subsubsection{Implementation Details}
\vspace{-0.2em}
We implemented the \textit{mixup}, MT, ICT, and our proposed MBT to train the state-of-the-art speech separation model -- Conv-TasNet \cite{luo2018conv} for comparative performance analysis. The Conv-Tasnet model architecture was replicated exactly from \cite{luo2018conv}. 
%For optimization, we used the built-in Adam optimizer \cite{kingma2014adam} in PyTorch to optimize the objective functions defined in each training method.
In all SSL settings including MT, ICT and our MBT method, we set the same decay coefficient for the mean-teacher to 0.999 to remain conservativeness following \cite{tarvainen2017mean}, and the same ramp function $r(t)=\exp(t/T_\text{max}-1)$ for $t\in\{1,...,T_\text{max}\}$, where $T_\text{max}=100$ was the maximum number of epochs. Besides, we set $\alpha=1$ following \cite{zhang2017mixup}, so that $\lambda$ becames uniformly distributed in $[0, 1]$.
\subsection{MBT for Supervised Learning}
\vspace{-0.3em}
% \begin{table}[h!]
% \centering
% \small
% \begin{tabular}{c|c|c}
% \hline
% \textbf{Method} & \textbf{SI-SNRi} & \textbf{SDRi} \\
% \hline
% Conv-TasNet w/o MBT & 15.3 & 15.6 \\
% Conv-TasNet w/ MBT & \bf{15.5} & \bf{15.9}\\\hline
% \end{tabular}
% \caption{Separation performance of Conv-TasNet on the WSJ0-2mix dataset with or without MBT}
% \label{tab:4}
% \end{table}

\begin{table}[h!]

\caption{\small Comparison of performances on the WSJ0-2mix dataset. $\star$: Results in our implementation.}
\centering
\small
\begin{tabular}{c|c|c|c}
\hline
{\textbf{Method}} & {\textbf{Params.}} & \textbf{Trained} & \textbf{SI-SNRi} \\
\hline
\scriptsize{DPCL++\cite{isik2016single}}   & 13.6M & & 10.8 \\
% \scriptsize{uPIT-BLSTM\cite{kolbaek2017multitalker}} & 92.7M & & - &  10.0 \\
% \scriptsize{cuPIT-Grid \cite{xu2018single}} & 47.2M & & - & 10.2 \\
\scriptsize{DANet \cite{chen2017deep}} & 9.1M & & 10.5 \\
\scriptsize{ADANet \cite{luo2018speaker}} & 9.1M &WSJ0-2mix& 10.4 \\
\scriptsize{Chimera++ \cite{wang2018alternative}} & 32.9M & & 11.5  \\
\scriptsize{WA-MISI-5 \cite{wang2018end}} & 32.9M & & 12.6 \\
\scriptsize{BLSTM-TasNet \cite{luo2018real}} & 23.6M & & 13.2 \\
\scriptsize{$^\star$Conv-TasNet} & 8.8M & & 15.3 \\\hline
 & &WSJ0-2mix+& \\
$^\star$MBT & 8.8M &``online" data& \bf{15.5} \\
 & &augmentation& \\\hline
 & &WSJ0-2mix+& \\
$^\star$MBT & 8.8M &Unlabeled& \bf{15.6} \\
 & &WSJ0-multi&  \\\hline
\end{tabular}
\vspace{-1em}
\label{tab:4}
\end{table}
First, we conducted a benchmark evaluation on the  WSJ0-2mix dataset without using any unlabeled data to examine the performance of MBT simply as an ``online" data augmentation for purely supervised learning. This learning was done by directly computing the consistency loss only on the labeled data, i.e., let $\mathcal{D}=\{\mathbf{x}_i\}_{i=1}^{N_L}$.
The separation performances of different supervised systems on the WSJ0-2mix dataset shows in Table \ref{tab:4}. Despite the very strong Conv-TasNet as baseline, by replacing the conventional ERM with MBT in Conv-TasNet, MBT managed to enhance the performance by $0.2$ and $0.3$ absolute SI-SNRi and SDRi improvement respectively.

\vspace{-0.5em}
\subsection{Generalization Capability}
\label{sec:generalization}
\vspace{-0.3em}
Since MBT suited both supervised learning and SSL framework, for training Conv-TasNet, we compared MBT with two sets of reference approaches: 1) supervised learning methods including the ERM and the \textit{mixup}, and 2) SSL methods including the MT and the ICT.

The goal in this section was to study and compare the generalization capability of different methods. For training the SSL models, we assume they have access to the corresponding unlabeled training sets of WSJ0-Libri, WSJ0-noise, and WSJ0-music. Each model was then evaluated on the test sets of WSJ0-Libri, WSJ0-noise, and WSJ0-music, representing different kinds of interference with increasing degrees of mismatch.

\vspace{-0.5em}
\subsubsection{Mismatch Speech Interference}
\vspace{-0.2em}
The first task evaluated two-talker speech separation with mismatch speech interference between training and inference. In the supervised learning setting, as shown in the upper half of Table \ref{tab:1}, MBT obtained higher SI-SNRi over the ERM and the \textit{mixup} baseline; in the SSL setting, it also outperformed the MT and the ICT.

\begin{table}[h!]
\vspace{-0.5em}
\caption{\small Separation performance of different training approaches in the presence of mismatch speech interference}
\centering
\small
\begin{tabular}{c|c|c|c}
\hline
\textbf{Method} & \textbf{Trained on} & \textbf{Tested on} & \textbf{SI-SNRi} \\
\hline
ERM  &  &  & 13.56 \\
mixup  & WSJ0-2mix & & 13.58 \\
MBT  & & & \bf{13.75} \\\cline{1-2} \cline{4-4}
MT &\multirow{2}{*}{WSJ0-2mix+}& WSJ0-Libri & 13.81 \\
ICT &\multirow{2}{*}{Unlabeled WSJ0-Libri}& & 13.78 \\
MBT & & & \bf{13.95}  \\\cline{1-2} \cline{4-4}
\multirow{2}{*}{MBT} &WSJ0-2mix+ & &\multirow{2}{*}{13.88}\\
 &Unlabeled WSJ0-multi& &  \\\hline

\end{tabular}
\vspace{-1em}
\label{tab:1}
\end{table}

\vspace{-0.5em}
\subsubsection{Mismatch Background Noise Interference}
\label{sec:noise}
\vspace{-0.2em}
Moreover, we tested the robustness of the separation models in the case of unseen background noise collected in real-world environments. We supposed this task might be less challenging than separating the two-talker speech mixture since background noise tends to be stationary, yet, unexpectedly, as shown in the upper half of Table \ref{tab:3}, all supervised learning systems failed to retain the performance of what has achieved in domain of WSJ0-2mix. The result reflects the semi-supervised learning is crucial for a robust separation system, as shown in the lower half of Table \ref{tab:3}, that it can achieve high and much more acceptable separation performance than the supervised systems in a new domain without any labeled data. Consistently, the MBT outperformed the MT and ICT by $0.7$ and $0.85$ absolute SI-SNRi improvement, respectively.
\begin{table}[h!]
\vspace{-0.5em}
\caption{\small Separation performance of different training approaches in the presence of mismatch background noise interference}
\centering
\small
\begin{tabular}{c|c|c|c}
\hline
\textbf{Method} & \textbf{Trained on} & \textbf{Tested on} & \textbf{SI-SNRi} \\
\hline
ERM  & \multirow{3}{*}{WSJ0-2mix} & & 1.86 \\
mixup  & & & 1.91 \\
MBT  & & & \bf{2.10} \\\cline{1-2} \cline{4-4}
MT & \multirow{2}{*}{WSJ0-2mix +} & \multirow{3}{*}{WSJ0-noise} & 12.51 \\
ICT & \multirow{2}{*}{Unlabeled WSJ0-noise} & & 12.36 \\
MBT & & & \bf{13.21}  \\\cline{1-2} \cline{4-4}
\multirow{2}{*}{MBT} & WSJ0-2mix + & &\multirow{2}{*}{\bf{13.52}} \\& Unlabeled WSJ0-multi & &  \\\hline
\end{tabular}
\vspace{-1em}
\label{tab:3}
\end{table}
\vspace{-0.5em}
\subsubsection{Mismatch Music Interference}
\vspace{-0.2em}
The third task was considered the most challenging since the music interference was highly non-stationary and was of a completely out-domain audio type. Similar to the result in Sec \ref{sec:noise}, the upper half in Table \ref{tab:2} indicates all supervised systems were drastically degraded to around $2$ SI-SNRi, whereas the systems trained with semi-supervised learning could remain high standard of performance, in which MBT produced significantly higher SI-SNRi than both MT and ICT with up to $13.77\%$ relative improvement.
\begin{table}[h!]
\vspace{-0.5em}
\caption{\small Separation performance of different training approaches in the presence of mismatch music interference}
\centering
\small
\begin{tabular}{c|c|c|c}
\hline
\textbf{Method} & \textbf{Trained on} & \textbf{Tested on} & \textbf{SI-SNRi} \\
\hline
ERM  & \multirow{3}{*}{WSJ0-2mix} & \multirow{3}{*}{ } & 1.93 \\
mixup  & & & 1.94 \\
MBT  & & & \bf{1.99} \\\cline{1-2} \cline{4-4}
MT & \multirow{2}{*}{WSJ0-2mix +} & \multirow{3}{*}{WSJ0-music} & 14.12 \\
ICT & \multirow{2}{*}{Unlabeled WSJ0-music} & & 14.02 \\
MBT & & & \bf{15.95} \\\cline{1-2} \cline{4-4}
\multirow{2}{*}{MBT} & WSJ0-2mix + & &\multirow{2}{*}{15.67} \\& Unlabeled WSJ0-multi & & \\\hline
\end{tabular}
\vspace{-1em}
\label{tab:2}
\end{table}

\vspace{-0.5em}
\section{Conclusions}
\vspace{-0.5em}
\label{sec:conc}
This paper introduces a novel training method called Mixup-Breakdown training (MBT). It can significantly improve the generalization of the state-of-the-art speech separation models over the standard training framework. The contribution of MBT is mainly two-fold: First, when given only labeled data, MBT can serve as a superior data augmentation technique for speech separation; secondly, when provided with a large amount of unlabeled speech mixture signals that possibly contain mismatch interference, MBT can effectively exploit the unlabeled data to enhance the generalization power of separation model with a minimal computational overhead. Our results indicate that MBT can remain strong performance even in non-specific domain. We will explore more perturbation types for MBT to further unleash its generalization capability in future work.


\small
% \linespread{0.9}
\bibliographystyle{IEEEbib}
\bibliography{strings,refs}

\end{document}